\documentclass{aa}
\usepackage{txfonts}

\usepackage{natbib}
\bibpunct[, ]{(}{)}{;}{a}{}{,}
\usepackage{graphicx}
\usepackage[english]{babel}
\usepackage{lscape}

\newcommand{\lppr}{\stackrel{<}{\scriptstyle \sim}}
\newcommand{\lappr}{\raisebox{-0.4ex}{$\lppr$}}

\def\spose#1{\hbox to 0pt{#1\hss}}
\def\lta{\mathrel{\spose{\lower 3pt\hbox{$\mathchar"218$}}
           \raise 2.0pt\hbox{$\mathchar"13C$}}}
\def\gta{\mathrel{\spose{\lower 3pt\hbox{$\mathchar"218$}}
           \raise 2.0pt\hbox{$\mathchar"13E$}}}
\newcommand{\Porb}{\mbox{$P_\mathrm{orb}$}}

\newcommand{\Mwd}{\mbox{$M_\mathrm{wd}$}}
\newcommand{\Msec}{\mbox{$M_\mathrm{sec}$}}

\newcommand{\Msun}{\mbox{$M_{\odot}$}}

\begin{document}

\title{The dwarf nova SS\,Cygni: what is wrong?}
\authorrunning{Schreiber \& Lasota}
\titlerunning{The dwarf nova SS\,Cygni: what is wrong?}
\author{Matthias R. Schreiber\inst{1}, Jean-Pierre Lasota\inst{2,3}}
\institute{
Departamento de F\'isica y Astronom\'ia, Facultad de Ciencias, Universidad
de Valpara\'iso, Valpara\'iso, Chile
\and Institut d'Astrophysique de Paris, UMR 7095 CNRS, Universit\'e
P. et M. Curie,
98bis Boulevard Arago, 75014 Paris, France
\and Astronomical Observatory, Jagiellonian University,
ul. Orla 171, 30-244 Krak\'ow, Poland\\
\email{Matthias.Schreiber@uv.cl; lasota@iap.fr} }

\offprints{M.R. Schreiber}

\date{Received / Accepted }

\abstract{Since the Fine Guiding Sensor (FGS) on the Hubble
Space Telescope (HST) was used to measure the distance to SS\,Cyg to
be $166\pm12$\,pc, it became apparent that at this distance the disc
instability model fails to explain the absolute magnitude during
outburst. It remained, however, an open question whether the model
or the distance have to be revised. Recent observations led to a
revision of the system parameters of SS\,Cyg and seem to be
consistent with a distance of $d\gta\,140$\,pc} {We re-discuss the
problem taking into account the new binary and stellar parameters
measured for SS\,Cyg. We confront not only the observations with the
predictions of the disc instability model but also compare SS\,Cyg
with other dwarf novae and nova-like systems. } {We assume the disc
during outburst to be in a quasi stationary state  and use the
black-body approximation to estimate the accretion rate during
outburst as a function of distance. Using published 
analysis of the long term light curve we determine the mean mass transfer 
rate of SS\,Cyg
as a function of distance and compare the result with mass transfer
rates derived for other dwarf novae and nova-like systems. } {At a
distance of $d\gta\,140$\,pc, both the accretion rate during
outburst as well as the mean mass transfer rate of SS\,Cyg
contradict the disc instability model. More important, at such
distances we find the mean mass transfer rate of SS\,Cyg to be
higher or comparable to those derived for nova-like systems.} {Our
findings show that a distance to SS\,Cyg $\gta 140$\,pc contradicts
the main concepts developed for accretion discs in cataclysmic
variables during the last 30 years. Either our current picture of
disc accretion in these systems must be revised or the distance to
SS\,Cyg is $\sim\,100$\,pc.}

\keywords{accretion, accretion discs -- instabilities --
stars: individual: SS\,Cyg -- stars: novae, cataclysmic variables --
stars: binaries: close}

\maketitle

\section{Introduction}
Dwarf novae are weakly magnetized cataclysmic variables (CVs)
showing quasi-regular outbursts, i.e., increased visual brightness
of 2-5\,mag for several days, which typically reappear on timescales
of weeks to months \citep[e.g.][for a review]{warner95-1}.

The standard disc instability model (DIM) assumes a constant
mass-transfer rate through the whole outburst cycle and is
successful in explaining the basic properties of dwarf nova
outbursts. In general the rise to maximum and the decay of
\textsl{normal} outbursts is well described by the standard
version of the model. There are problems with quiescence.
Superoutbursts, Z\,Cam-type outbursts, and the reproduction of the
outburst cycle in general require DIM modifications \citep[see][for
a review]{lasota01-1}.

On the other hand the DIM is too simple to be a faithful
representation of dwarf nova outbursts. It uses a 1+1D scheme and is
based on the $\alpha$--parameter description of viscosity. Although
the news of the death of such an approach \citep{pessahetal06-1} are
exaggerated, its serious limitation have been well known for a
long time.

It is therefore not surprising that the brightest and best observed
dwarf nova SS Cyg has been a source problems for the DIM. Its
various types of outbursts seem to require modulations of the
mass-transfer rate and the anomalous outbursts remain unexplained
\citep[e.g.][]{schreiberetal03-1}. However, the main challenge
comes from the distance to this system obtained from the HST/FGS
parallax \citep{harrisonetal99-1}. According to the DIM, at
such a distance ($166\pm12$\,pc) the accretion disc of SS Cyg would
be hot and stable and the system would not be a dwarf nova.
Therefore such a distance, if correct, would seriously put in doubt
the validity of the DIM. Indeed, \citet{schreiber+gaensicke02-1}
concluded that either the DIM has to be modified and strongly
enhanced mass transfer during outbursts plays an important
role, or the distance of $166\pm12$pc is wrong. Comparing
detailed DIM simulations with the observations of SS\,Cyg,
\citet{schreiberetal03-1} assumed $d=100$\,pc doubting the
correctness of higher values.

Recently \citep{bitneretal07-1} observationally re-determined
the parameters of SS\,Cyg and obtained values for the masses of the
stellar components and the orbital inclination that differ
significantly from those derived earlier. The new results are more
reliable than earlier measurements because they do not rely on
error-prone methods such as those based on the wings of emission
lines to determine the mass ratio or those using a main sequence
mass/radius relation to derive the orbital inclination
\citep[see][for a detailed discussion]{bitneretal07-1}. Very
important in the context of the DIM is the conclusion of
\citet{bitneretal07-1} that their results are consistent 
with a distance of $d\sim$ 140 -- 170 pc in line with the parallax 
measurement.
This forced us to re-examine the problem.

The structure of the paper is as follows. In Sect. \ref{odl}
applying the method of \citet{schreiber+gaensicke02-1}
but using the revised system
parameters of \citet{bitneretal07-1} we
compare the predicted absolute magnitude and the accretion
rate during normal outbursts with
the value derived from observations. Thereafter we determine the
mean mass-transfer rate from the observed outburst properties
(Sect.\ref{mtr}) and again compare it with the predictions of the DIM.
The conclusion of these two investigations
is that a distance of $166\,\pm12$pc is incompatible
with the DIM. In Sect. \ref{ssnl} we
compare the mean mass-transfer rates of SS\,Cyg and other dwarf
novae with HST/FGS-parallax measurements with those of several nova-like
binaries. We show that at the HST/FGS parallax distance
SS\,Cyg is \textsl{brighter} than some nova-like systems
and conclude that being an outbursting system at such high
luminosity SS\,Cyg must be a very special CV indeed. In the
following (Sect.\,\ref{emt}) we re-consider the possibility
that in SS\,Cyg the mass-transfer rate increases during outbursts as
this would lower the mean mass-transfer rate.

\section{Accretion rate during outburst}
\label{odl}

\begin{figure}
\includegraphics[width=6.5cm, angle=270]{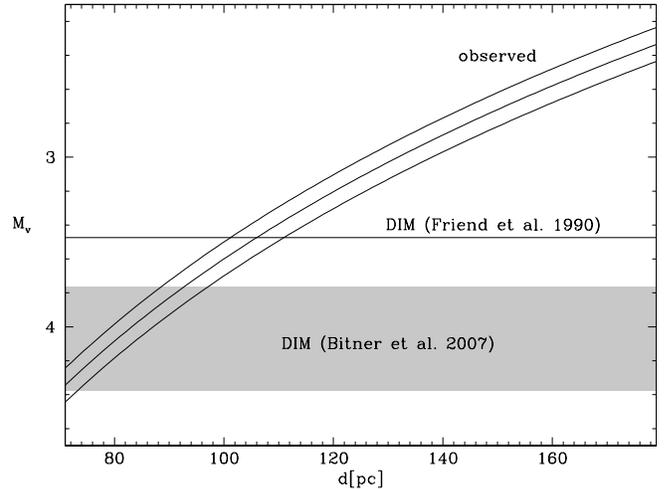}
\caption{\label{fond} The absolute magnitude at the onset of the
decline derived from the observed magnitude of $m_v=8.6\,\pm0.1$ as
a function of distance and the absolute magnitude predicted by the
DIM according to system parameters derived by \citet{bitneretal07-1}
and \citet{friendetal90-1} (see Table\,1). }
\end{figure}

One of the key-predictions of the DIM is that at the onset of the
decline, i.e. when the cooling front forms at the
outer edge of the disc, the disc is in a quasi-stationary outburst
state and the mass accretion rate is close to the critical mass
transfer rate given by
\begin{equation}
\dot{M}_{\mathrm{crit}}=9.5\times\,10^{15}\,\mathrm{g\,s^{-1}}\,R_{10}^{2.68}\,M_{\rm wd}^{-0.89},
\end{equation}
where $\Mwd$ is the white-dwarf's mass is solar units and $R_{10}$
the disc radius in units of $10^{10}$ cm \citep[e.g.][]{hameuryetal98-1}.
The light curves of normal outbursts of SS\,Cyg show a plateau phase with
nearly constant brightness before the onset of the decline.
Therefore, according to the DIM, the accretion rate during outburst should be
similar to the critical accretion rate.
Assuming the outer radius of the disc to be close to the tidal truncation
radius
$R_{\mathrm{out}}=0.9\,R_1$ with $R_1$ being the primary's Roche-lobe radius,
we derive for the new parameters of SS\,Cyg (see Table\,1)
\begin{equation}
\dot{M}_{\mathrm{crit}}\sim\,9.0-9.1\times10^{17}\mathrm{g/s}.
\end{equation}
Clearly, we can derive the predicted absolute visual magnitude of a disc with
this accretion rate. We follow \citet[][]{schreiber+gaensicke02-1}, i.e. we
use the same equation to account for the inclination, assume the
effective temperature to follow the radial dependence of stationary
accretion discs and the annuli of the disc to radiate like black
bodies. For the mass accretion rate given in Eq.\,(2) we than obtain
$M_V=3.76-4.37$ as the predicted absolute magnitude during outburst.
On the other hand, using the observed visual magnitude, i.e. $m_V=8.6\pm0.1$,
we can determine the absolute magnitude as a function of distance.
Figure\,\ref{fond} compares both values. The shaded region
represents the absolute magnitude predicted by the DIM for the range
of system parameters for SS\,Cyg recently derived by
\citet{bitneretal07-1} summarized in Table\,1.
Also shown in Fig.\,\ref{fond} (solid horizontal line)
is the predicted absolute magnitude when using the same system parameter as
\citet{schreiber+gaensicke02-1}, i.e. those given by
\citet{ritter+kolb98-1} which are based on \citet{friendetal90-1}.
Apparently, the range of predicted absolute magnitudes significantly
decreased due to the change of the system parameters. As a
consequence of the revised inclination and mass of the white dwarf
(see Table\,1), agreement with the DIM now requires distances as
short as $d\lappr\,100$\,pc.

\begin{table}
\begin{center}
\caption[]{\label{t-ssc_par} Binary parameters of SS\,Cyg used by
\citet{schreiber+gaensicke02-1} (based on
\citep{friendetal90-1} and \citep{ritter+kolb98-1})
and the new values recently derived by
\citet{bitneretal07-1} (right column). Also given is the derived
outer radius of the disc ($\sim0.9R_1$) in units of $10^{10}$\,cm. 
Please note that the stellar masses (and therefore also $R_{10}$) can not be 
chosen independently from
within their ranges as they are strongly constrained by the mass ratio.
}
%
% Tabelle etwas enger setzen
\setlength{\tabcolsep}{2ex}
\begin{tabular}{lcr}
\hline\hline\noalign{\smallskip}
& Ritter \& Kolb (1998) & Bitner et al. (2007) \\
\hline\noalign{\smallskip}
\Porb/hr       & 6.6  &  6.6 \\
$\Mwd/\Msun$ & $1.19\pm0.05$ & $0.81\pm0.19$ \\
$q=\Msec/\Mwd$ & 0.70 & $0.683\pm0.012$\\
$i/^o$ & 37 & 45-56 \\
$R_{\mathrm{10}}$ & $5.73$ & $4.7-5.5$ \\
%-----------------------------------------------------------------
\noalign{\smallskip}\hline\noalign{\smallskip}
\end{tabular}
\linebreak
\end{center}
\end{table}

To see how the discrepancy in the absolute magnitudes
for a distance of $166\pm\,12$\,pc correlates with accretion rates,
we also compare the value of the critical accretion
rate (Eq.\,(2)) with the accretion rate required to reproduce the
absolute magnitude assuming SS\,Cyg is at $d=166\pm12$\,pc.
We find that an accretion rate of
\begin{equation}
\dot{M}_{\mathrm{out}}\sim\,8.8-9.2\times\,10^{18}\mathrm{g/s}
\end{equation}
is required. This is {\em{an order of magnitude}}
above the value predicted by the DIM (Eq.\,(2)).
The value given above is also essentially higher (by a
factor of $\sim\,2.5$) than the
one derived by \citet{schreiber+gaensicke02-1}. The higher mass accretion
rate is required because \citet{bitneretal07-1} found a higher value for the
inclination and a lower value for the mass of the white dwarf.

\section{The mean mass transfer rate}
\label{mtr}

SS\,Cyg is among the visually brightest dwarf nova and a detailed long-term
light curve exists. The mean outburst properties have been derived by
\citet{cannizzo+mattei92-1,cannizzo+mattei98-1} who analysed the AAVSO long
term light curve.
They find a mean outburst duration of $t_{\mathrm{out}}=10.76$ days, a mean
cycle duration of $t_{\mathrm{cyc}}=49.47$\,d giving a mean quiescence time
of $t_{\mathrm{qui}}=38.71$\,d. The mean duration of rise to outburst and
decline from outburst are $t_{\mathrm{ris}}=0.5$\,d and
$t_{\mathrm{dec}}=2.5$\,d respectively.
Using these values we can now derive a value for the mean mass transfer
rate from the observed visual magnitude during outburst.
As in Sect.\,\ref{odl} we will compare the value derived from the observations
with the prediction of the DIM.

According to the DIM, the mean mass-transfer rate can be obtained from the
relation
\begin{equation}
t_{\rm qui}\approx \frac{\epsilon M_{\rm D, max}}{\dot M_{\rm tr}-
\dot M_{\rm in}}
\label{treccu}
\end{equation}
where the fraction of the disc's mass lost during outburst is $\epsilon=
\Delta M_{\rm D}/M_{\rm D, max}$ and  $\dot M_{\rm in}$ is the accretion
rate at the disc's inner edge. Usually ${\dot M_{\rm tr}>>
\dot M_{\rm in}}$ and $\epsilon\sim0.1$. Taking
\begin{equation}
M_{\rm D, max}=2.7 \times 10^{21} \alpha^{-0.83} M_{\rm wd}^{-0.38} R_{10}^{3.14} \ {\rm g},
\label{diskmass}
\end{equation}
\citep[see][]{lasota01-1}, $\alpha=0.02$, and the system parameters
derived by \citet{bitneretal07-1} we obtain
\begin{equation}
\dot{M}_{\rm{tr}}=2.6-4.2\times\,10^{17}{\rm{g/s}}.
\label{mtrdim}
\end{equation}

The above value should be compared with the mean mass transfer rate
derived from the observed visual brightness during outburst.
Following again \citet[][their Eq.\,(5)]{schreiber+gaensicke02-1}
but using the system parameters obtained by \citet{bitneretal07-1},
we derive a mean mass transfer rate as function of distance. For
$d=166\,$pc we obtain
\begin{equation}
\dot{M}_{\rm{tr}}=1.1-3.8\times\,10^{18}{\rm{g/s}}.
\label{mtrobs}
\end{equation}

\begin{figure}
\includegraphics[width=6.5cm, angle=270]{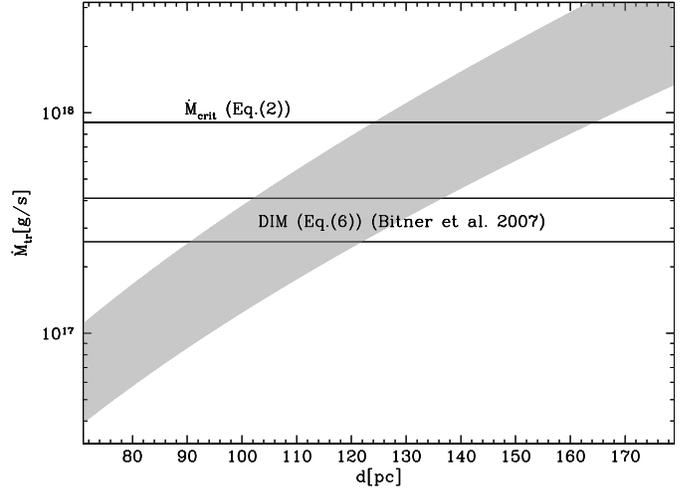}
\caption{\label{fmean} The mean mass transfer rate of SS\,Cyg
derived from the observed visual magnitude as a function of distance
(shaded region) compared with the predictions of the DIM (horizontal
lines). All values are calculated for the range of system parameters
derived by \citet{bitneretal07-1}. Agreement between DIM and
observed visual magnitude requires a distance of $d\sim100\,$pc. At
a distance of $166$pc the mean mass transfer derived from the
observations is above the critical transfer rate and - according to
the DIM - SS\,Cyg should be a nova-like.}
\end{figure}

The two values for the mean transfer rate, i.e. the one predicted by the DIM
and the one derived from the observations
are compared in Fig.\,\ref{fmean}.
The grey shaded region represents the values required by
$m_V=8.6\pm0.1$ as a function of distance.
The range of mean mass transfer rates predicted by the
DIM (Eq.\,(\ref{mtrdim})) and the critical mass transfer rate (Eq.\,(2)) are
shown as horizontal lines.
Again, the discrepancy between DIM and HST/FGS parallax is obvious:
According to the DIM, at $166\pm12$\,pc SS\,Cyg should be nova-like
system and not a dwarf nova. Even for $d\sim\,140$\,pc, the derived
mean mass transfer rates are close to the critical value and one
would at least expect Z\,Cam-like behaviour. Again,
agreement with the DIM requires a distance of $d\sim100$\,pc.

Although the recently determined parameters significantly increase
the discrepancy between HST/FGS parallax and DIM prediction, the
problem has been mentioned and discussed earlier.
\citet{schreiber+gaensicke02-1} proposed as one possible solution a
revision of the DIM by assuming an increased value of the critical
mass transfer rate which would be equivalent to allowing for dwarf
nova outbursts for higher mass transfer rates. However, as we will
see in the next section the problem is not with the DIM. At
$166\pm12$\,pc the mean mass-transfer rate of SS\,Cyg is comparable
or higher than that of nova-like binaries with similar orbital
parameters but unlike SS\,Cyg these systems never show outbursts. If
anything they show a so-called ``anti-dwarf-nova" behaviour.

\section{Comparing SS\,Cyg with nova-like CVs}
\label{ssnl}

\begin{figure}
\includegraphics[width=6.5cm, angle=270]{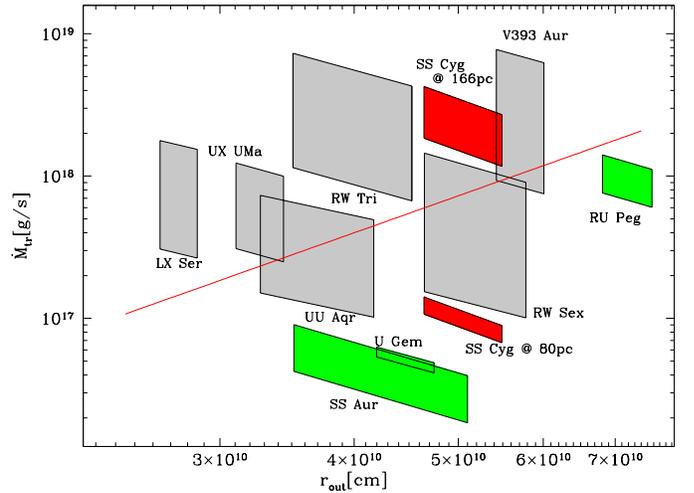}
\caption{\label{fig_mtr_all} The mean mass transfer rate of SS\,Cyg,
three other dwarf novae with HST/FGS distance, and six well known
nova-like CVs as a function of the outer radius of the disc during
outburst. As binary parameters of SS\,Cyg  we used again the values
(and uncertainties) derived by \citet{bitneretal07-1}. The
parameters and distances used for the nova-likes are compiled and
discussed in Table\,1. Both, in order to make the plot easier to
read as well as because the broad ranges of possible parameters do
not represent well-determined values with certain errors, we use
shaded boxes instead of error bars. The solid line represents the
critical mass transfer rate according to Eq.\,(1) assuming
$\Mwd=1\Msun$. According to the DIM, this line should separate dwarf
novae and nova-like systems. Interestingly, to reach agreement with
this prediction for SS\,Cyg at $d=166$\,pc, the size of the disc
needs to be similar to the one in RU\,Peg ($\sim7\times10^{10}$cm).
Even for the maximum mass of the white dwarf
($\Mwd=1.4\Msun$) this would require a disc larger than the Roche-lobe radius
of the primary ($6.8\times10^{10}$\,cm). }
\end{figure}

\begin{table*}
\newcounter{ref}
\newcommand{\tcite}{\stepcounter{ref}\arabic{ref}}
\newcommand{\tref}[1]{\stepcounter{ref}(\arabic{ref})\,\citealt{#1}}
%\begin{center}
\caption[]{\label{tnl} Binary parameters, distances, visual
magnitudes, and extinctions of $6$ nova-like systems.}
%
% Tabelle etwas enger setzen
\setlength{\tabcolsep}{3.5ex}
\begin{tabular}{lcccccccr}
\hline\hline\noalign{\smallskip}
name & \Mwd & \Msec & \Porb & d & i & $m_V$ & $A_V$ & ref. \\
\hline\noalign{\smallskip}
RW\,Tri & 0.4-0.7 &  0.3-0.4 & 5.565 & 310-370 & 67-80 & 13.2 & 0.3-0.7 & 1, 2, 3,
4, 5 \\
UU\,Aqr & 0.6-0.9 & 0.2-0.4 & 3.92 & 250-350 & 76-80 & 13.6 & 0-0.2 & 5, 6 \\
LX\,Ser & 0.37-0.43 & 0.3-0.4 & 3.80 & 300-400 & 77-83 & 14.4 & 0-0.2 & 4, 5\\
RW\,Sex & 0.8-1.3 & 0.55-0.65 & 5.88 & 150-250 & 30-40 & 10.8 & 0-0.2 & 5, 7\\
UX\,UMa & 0.4-0.5 & 0.4-0.5 & 4.72 & 200-300 & 69-73 & 12.8 & 0-0.2 & 4, 5 \\
V363\,Aur & 0.8-1.0 & 0.8-1.0 & 7.71 & 600-1000 & 68-72 & 14.2 & 0.3-0.5 & 4, 8 \\
%-----------------------------------------------------------------
\noalign{\smallskip}\hline\noalign{\smallskip}
\end{tabular}
\linebreak For some systems the values given in the literature
differ significantly. To keep our results as independent as possible
of uncertainties related to the system parameters of nova-likes, we
always used a broad range of parameters. The values of $A_V$ have
been taken from \citet{bruch+engel94-1} and compared with
\citet{warner87-1} who quotes \citet{bruch84-1}. \setcounter{ref}{0}
References:
% RW Tri
(1) \citet{mcarthuretal99-1}, (2) \citet{pooleetal03-1}, (3)
\citet{grootetal04-1}, (4) \citet[][and references therein]{ruttenetal92-1},
(5) \citet{vandeputteetal03-1}
% UU Aqr
(6) \citet{baptistaetal94-1},
%LX Ser
%RW Sex
(7) \citet{beuermannetal92-1},
% UX UMa
% V363 Aur
(8) \citet{thoroughgoodetal04-1}
%\end{center}
\end{table*}

\begin{table*}
\newcounter{ref2}
\newcommand{\tcite}{\stepcounter{ref}\arabic{ref2}}
\newcommand{\tref}[1]{\stepcounter{ref}(\arabic{ref2})\,\citealt{#1}}
\caption[]{\label{t-dn_par} Binary and light curve parameters of the
four dwarf novae with HST/FGS parallax. The time the disc is in the
quasi-stationary state during outburst $t_{\rm{qs}}$ is approximated as in
\citet{schreiber+gaensicke02-1}. }
%
% Tabelle etwas enger setzen
\setlength{\tabcolsep}{3ex}
\begin{tabular}{lccccccccr}
\hline\hline\noalign{\smallskip}
name & \Mwd & \Msec & \Porb & d & i & $m_V(out)$ & $A_V$ & $t_{\mathrm{qs}}/t_{\mathrm{cyc}}$ & ref. \\
\hline\noalign{\smallskip}
SS\,Cyg & 0.6-1.0 & 0.4-0.7 & 6.6 & $166\pm12$ & 45-56 & 8.5 & 0-0.2 & 5.8/49.5 & 1, 2, 3, 4 \\
U\,Gem & 1.0-1.3 & 0.45-0.5 & 4.25 & 90-100 & 67-71 & 9.3-9.6 & 0-0.1 & 5/118
& 4, 5, 6 \\
SS\,Aur & 0.6-1.4 & 0.38-0.42 & 4.39 & 175-225 & 32-47 & 10.7-10.9 & 0.1-0.3 & 5/53 & 4, 5 \\
RU\,Peg & 1.1-1.4 & 0.9-1.0 & 8.99 & 261-303 & 34-48 & 9.0-9.1 & 0-0.1 & 6/75 & 4, 7 \\
%-----------------------------------------------------------------
\noalign{\smallskip}\hline\noalign{\smallskip}
\end{tabular}
\linebreak Again, we used a rather broad range of parameters in
order to avoid our conclusions depending on uncertain parameters.
For completeness we added again the parameters ranges for SS\,Cyg
according to \citet{bitneretal07-1}. Please note that for SS\,Cyg
$\Mwd$ and $\Msec$ are constrained by $q=0.683\pm0.012$.
\setcounter{ref}{0} References: (1) \citet{bitneretal07-1}, (2)
\citet{harrisonetal99-1}, (3) \citet{cannizzo+mattei92-1}, (4)
\citet{harrisonetal04-1}, (5) \citet{szkody+mattei84-1} (6)
\citet{nayloretal05-1} (7) \citet{aketal02-1}
\end{table*}

There is an overwhelming evidence that the accretion disc is the site of
dwarf-nova outbursts.
The general picture of disc accretion in CVs is that below a certain mass
transfer rate the disc is unstable and dwarf nova outbursts occur. For higher
mass transfer rates, the disc is stable and the corresponding class of
CVs are nova-like systems. In agreement with this picture,
the mean absolute magnitudes of dwarf novae have been found to be lower than
those of nova like systems \citep[see][Fig.\,9.8]{warner95-1}.
To check whether this agreement remains for a distance
to SS\,Cyg of $166\pm12$\,pc, we compare the mean mass transfer rate
derived for SS\,Cyg with those obtained
for a set of well observed nova-like systems and three additional dwarf nova
with measured HST/FGS parallax (see Table\,\ref{tnl} and \ref{t-dn_par}).

Figure\,\ref{fig_mtr_all} \citep[inspired by Fig 1 in][]{smak83-1}
shows the derived mean mass transfer rates as a function of the
outer radius of the disc during outburst. To avoid our results
depending on uncertainties in the system parameters derived from
observation we used rather broad ranges of parameters. The straight
line represents the critical mass transfer rate for $\Mwd=1\Msun$.
Obviously, at a distance of $166$\,pc the mean mass transfer rate of
SS\,Cyg is claerly above this limit as discussed earlier (Fig.\,2).
The other three dwarf novae are below the dividing line and the
nova-likes have mass transfer rates higher than (or similar to) the
critical rate. The striking point of Fig.\,\ref{fig_mtr_all} is the
fact that the mean mass transfer rate of SS\,Cyg is larger (or as
large) as those derived for nova-like systems with similar system
parameters. In other words, if SS\,Cyg is indeed $d\gta140$\,pc
away, the difference between nova-like systems and the dwarf nova
SS\,Cyg cannot be in the mean mass transfer rate. This conclusions
represents a very important finding because it contradicts the
generally accepted picture for accretion discs in CVs.

Clearly, one could
argue that the distances to the nova-like systems might be systematically too
small. However, the distance to RW\,Tri is based on a HST/FGS parallax
and for the other systems we used very large upper limits for the
distance. Hence, there is no easy way out of the problem.
In the next section we will discuss a substantial revision of the DIM
that might provide a solution.

\section{Enhanced mass-transfer rate}
\label{emt}

\citet{smak00-1}, \citet{lasota01-1}, and \citet{smak05-1} showed that enhanced
mass transfer during outburst is
required to explain the light curve of U\,Gem, especially the extremely long
superoutburst in 1985.
Moreover, modulations of the mass-transfer rate are necessary to explain
outburst properties of SS\,Cyg itself \citep{schreiberetal03-1} and it seems that there is growing
evidence for enhanced mass transfer playing an important role in short orbital period dwarf novae of the
SU\,UMa type \citep[][]{schreiberetal04-1,smak04-3,smak05-1,sterkenetal07-1}.

The mean mass-transfer rate is
not an observed quantity but is calculated assuming constant
mass-transfer rate over the cycle, hence, in case the mass transfer rate is
significantly enhanced during outburst, our values are only upper limits.
In addition, in the framework of the enhanced mass transfer scenario, the accretion rate during
outburst and at the onset of the decline
does not need to be the critical mass-accretion rate. Therefore, if
evidence for a distance above $d=140\,$pc further grows, the enhanced mass
transfer scenario might be considered a possible solution.
The enhancement needed to put SS Cyg in the observed
dwarf nova band is quite substantial.
Assuming a mean mass transfer rate of $M_{\mathrm{tr}}=1.5\times10^{17}$g/s
during quiescence, \citet{schreiber+gaensicke02-1} estimated the required mass
transfer enhancement to be by about a factor of $\gta\,15$.

Taking into account the revisions of the system parameters according
to \citet{bitneretal07-1}, the required mass transfer enhancement
reaches a factor of $\gta\,55$ for $d=166$pc. Even at a distance of
$d=140$\,pc a factor of $\sim\,35$ is required. Compared to mass
transfer enhancements predicted for U\,Gem (factor 20-50,
\citet{smak05-1}) or SU\,UMa superoutbursts (15-60,
\citet[][]{smak04-3}) this seems to be plausible but one should keep
in mind the model calculations by \citet{smak04-1} which seem to
exclude irradiation induced enhancement for $P_{\rm orb}\gta\,6$\,h.
However, if a distance to SS\,Cyg of $140-170$\,pc will be further
confirmed in the future, considering enhanced mass transfer even in
(some) long orbital period dwarf novae appears to be the most
plausible mechanism to explain the observations.

\section{Conclusions}

The long term light curve of SS\,Cyg has been frequently used to constrain the disc instability model, in
particular the viscosity parameter $\alpha$.
Now, it seems that we can learn something very different but equally essential about accretion discs in CVs
from analysing this particular system.
SS\,Cyg is a dwarf nova and not a nova-like. It seems that the distance to SS\,Cyg is above $140$\,pc. If this
will be further confirmed,
then there is something we do not understand in this binary. The standard interpretation of mean mass transfer
rates that are constant over the outburst cycle cannot be true and a
difference in the mean mass transfer rate
cannot be the only difference between nova-likes and (at least one) dwarf
nova.
This might mean that the standard DIM is in fact not adequate and has to be
modified by including mass-transfer modulations. This is not a
surprise to these authors
\citep[e.g.][]{schreiberetal00-1,lasota01-1,buat-menardetal01-2,
schreiberetal04-1}.

\begin{acknowledgements}
JPL is grateful to Rob Robinson for helpful comments on observations
of SS Cyg.
MRS acknowledges support from FONDECYT (grant 1061199), DIPUV (project 35),
and the Center of Astrophysics in Valparaiso.
This research was supported in part by the
National Science Foundation under Grant No. PHY05-51164, 
report number: NSF-KITP-07-151.
\end{acknowledgements}

%\bibliography{../aamnem99,../aabib}
%\bibliographystyle{../apj}

%\begin{thebibliography}{30}
%\expandafter\ifx\csname natexlab\endcsname\relax\def\natexlab#1{#1}\fi

%\bibliography{../aamnem99,../aabib}
%\bibliographystyle{../apj}

\end{document}